\documentclass[default,iicol]{sn-jnl}

\usepackage{graphicx}%
\usepackage{multirow}%
\usepackage{amsmath,amssymb,amsfonts}%
\usepackage{amsthm}%
\usepackage{mathrsfs}%
\usepackage[title]{appendix}%
\usepackage{xcolor}%
\usepackage[final]{changes}%
\usepackage{textcomp}%
\usepackage{manyfoot}%
\usepackage{booktabs}%
\usepackage{algorithm}%
\usepackage{algorithmicx}%
\usepackage{algpseudocode}%
\usepackage{listings}%

\usepackage{braket}
\usepackage[T1]{fontenc}

\setaddedmarkup{\textcolor{red}{#1}}       
\setdeletedmarkup{\sout{#1}}               

\begin{document}


\title[Transformer Models for Quantum Gate Set Tomography]{Transformer Models for Quantum Gate Set Tomography}


\author*{\fnm{King Yiu} \sur{Yu}}\email{k.y.yu@tudelft.nl}

\author{\fnm{Aritra} \sur{Sarkar}}\email{a.sarkar-3@tudelft.nl}

\author{\fnm{Maximilian} \sur{Rimbach-Russ}}\email{m.f.russ@tudelft.nl}

\author{\fnm{Ryoichi} \sur{Ishihara}}\email{r.ishihara@tudelft.nl}

\author{\fnm{Sebastian} \sur{Feld}}\email{s.feld@tudelft.nl}


\affil{\orgdiv{Department of Quantum \& Computer Engineering}, \orgname{Delft University of Technology}, \orgaddress{\street{Mekelweg 5}, \city{Delft}, \postcode{2628 CD}, \state{Holland-Zuid}, \country{The Netherlands}}}
\affil{\orgdiv{QuTech and Kavli Institute of Nanoscience}, 
       \orgname{Delft University of Technology}, 
       \orgaddress{\city{Delft}, \country{The Netherlands}}}

\abstract{Quantum computation represents a promising frontier in the domain of high-performance computing, blending quantum information theory with practical applications to overcome the limitations of classical computation. This study investigates the challenges of manufacturing high-fidelity and scalable quantum processors. Quantum gate set tomography (QGST) is a critical method for characterizing quantum processors and understanding their operational capabilities and limitations. This paper introduces \textsc{Ml4Qgst} as a novel approach to QGST by integrating machine learning techniques, specifically utilizing a transformer neural network model. Adapting the transformer model for QGST addresses the computational complexity of modeling quantum systems. Advanced training strategies, including data grouping and curriculum learning, are employed to enhance model performance, demonstrating significant congruence with ground-truth values. We benchmark this training pipeline on the constructed learning model, to successfully perform QGST for \added{2 and 3 gates on single-qubit and two-qubit systems}, with over-rotation error and depolarizing noise estimation with comparable accuracy to pyGSTi.
This research marks a pioneering step in applying deep neural networks to the complex problem of quantum gate set tomography, showcasing the potential of machine learning to tackle nonlinear tomography challenges in quantum computing.}


\maketitle

\section{Introduction} \label{s1}
Quantum computation is an emerging paradigm of computation that has captured the attention of theoretical physicists and computer scientists, as well as stakeholders in high-performance computing.
Quantum algorithms can solve problems in specific complexity classes that are asymptotically intractable in all implementations of classical computation.
To demonstrate this acceleration in practice, a topical research avenue is on maturing the quantum computing hardware in terms of high-fidelity (decoherence, error rates of quantum operations) and scalability (number of qubits, connectivity) along with quantum software~\cite{soeken2018programming}, \cite{sarkar2024automated}.
Though this research endeavor has proved rather a challenging engineering feat~\cite{preskill2018quantum,leymann2020bitter,ezratty2023we}, rapid strides were made in the last decade with a plethora of physical technologies capable of demonstrating controllable processing of quantum information.

Constructing a quantum computer requires that the experimental setup meets certain conditions, succinctly summarized as the DiVincenzo criteria~\cite{divincenzo2000physical}.
A critical criterion is the characterization of the quantum processor, which helps in understanding the fabrication defects and the computing capabilities of these systems. 
The characterization is performed by building a model of the noise on the system and the set of quantum operations performed on the quantum information units.
The latter typically involves inferring the information-theoretic operation experimentally achieved with respect to a set of target gates in a circuit model quantum computer.
Once this operation -- termed quantum gate set tomography (QGST)~\cite{greenbaum2015introduction},~\cite{rudinger2021experimental} -- is performed, the updated experimental model of the quantum gates can be used precisely to determine the required sequence of quantum gates to achieve the transformation for a quantum algorithm.

Tomographic approaches build a detailed model for a system or component in a latent space by fitting that model to the data from numerous independent tests that reveal partial information about the system.  
The nature of this latent space depends on the representation of the system model.
For instance, in medical imaging, by aggregating information from multiple 2D sinograms of CT scans, one can reconstruct a full 2D/3D CT image.
Reconstructing a maximally fitting model for the experimental data is computationally expensive.
Quantum tomography is particularly expensive as the space of possible observations is continuous and grows exponentially with the system size.

Various machine learning techniques are used to address the computational cost in tomography~\cite{wang2019machine}.
However, these techniques have not yet been utilized for quantum gate set tomography.
Our contribution presented in this work is three-fold.
\textit{Firstly,} we develop a first-of-its-kind machine-learning model for quantum gate set tomography~(\textsc{Ml4Qgst}).
To this purpose, we harness the transformer neural network model~\cite{vaswani2017attention} and tune it for the input-output settings of QGST.
Generative models like GAN have been used for quantum state tomography~\cite{ahmed_quantum_2021}. 
However, QGST is substantially different due to non-linearity, multi-model regression, and worse computational scaling costs, making it challenging to reuse the models from other quantum tomography tasks.
As we show later, QGST can be framed as a language learning task, thus prompting our choice of employing the Transformer model. 
\textit{Secondly,} contrary to directly estimating the full process matrices in traditional algorithms, our approach ensures the resulting process matrices are always completely positive and trace preserving (CPTP) and without the necessity to perform gauge fixing.
Thus, instead of fully reconstructing the process matrices with no prior knowledge, we are interested in the pragmatic setting of inferring the model drift between the intended theoretical process and the experimentally achieved process.
And, \textit{thirdly}, we incorporate advanced training techniques of data grouping, curriculum learning, and computation of the loss function to increase the performance of our model.
Our model predicts error parameters from the error channels and subsequently generates the corresponding estimated process matrices of the gate set. 

\added{The article is structured as follows: Section~\ref{s2} introduces the problem setting and necessary definitions, while also contrasting this research with related works on machine learning for quantum tomography. Section~\ref{s3} provides a detailed explanation of the proposed method and the transformer architecture employed in this study. Advanced training techniques and experimental settings are discussed in Section~\ref{s4}. Section~\ref{s5} presents the results of the single-qubit experiments, and Section~\ref{2-qubit-intro} extends our approach to multi-qubit scenarios, followed by an analysis of the corresponding two-qubit experimental results in Section~\ref{2-qubit-experiment}. Finally, Section~\ref{Outlook} explores the future prospects of transformer-based neural network models for quantum tomography, and Section~\ref{s6} concludes the article.}

\section{Problem setting and definitions} \label{s2}

In this section, we present some required background for the article.
Firstly, the background theory of gate set tomography based on the super operator formalism is presented.
After that, we present how artificial neural networks can be employed for quantum tomography, directing the discussion towards the transformer model that is used in this research.

\subsection{Quantum gate set tomography} 

\begin{figure}
    \centering
    \includegraphics[width=\linewidth]{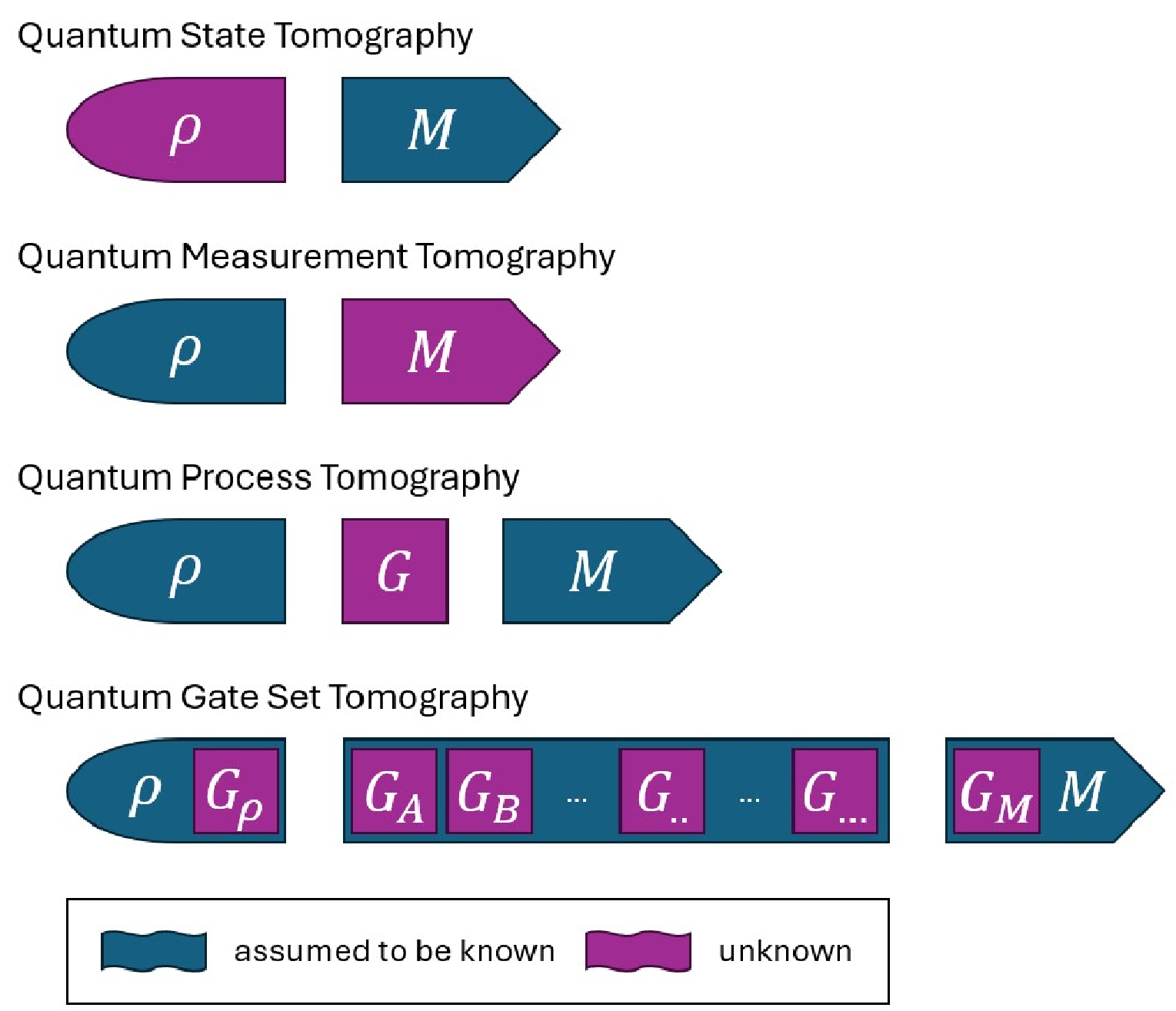}
    \caption{\added{The figure illustrates various quantum tomography protocols used to characterize quantum systems and processes. \textbf{Quantum State Tomography (QST):} The goal is to reconstruct the quantum state $\rho$ of a system by applying a set of measurements $M$ (assumed to be known) to the system. \textbf{Quantum Measurement Tomography (QMT):} In this case, the measurement operators $M$ are unknown, while the quantum state $\rho$ is assumed to be known. By preparing known quantum states, the measurements can be characterized. \textbf{Quantum Process Tomography (QPT):} Here, the quantum state $\rho$ and the measurements $M$ are assumed to be known, while the quantum process $G$ is unknown and has to be reconstructed. \textbf{Quantum Gate Set Tomography (QGST):} This approach generalizes earlier techniques by simultaneously reconstructing quantum gates $G$. State preparation $G_\rho$ and measurement $G_M$ are considered to be included in the full gate set that has to be characterized. Blue blocks indicate elements assumed to be known, while pink blocks represent unknown components to be reconstructed.}}
    \label{fig:qgst}
\end{figure}

Different from traditional quantum process tomography, which implicitly assumes known, hence near-zero state preparation and measurement (SPAM) errors, as shown in Figure~\ref{fig:qgst}, 
quantum gate set tomography relaxes
this assumption by directly incorporating gates as both preparation and measurement operators or formally as preparation and measurement fiducials. 
In a quantum computer characterization setting, rather than probing each individual gate using traditional process tomography, gate set tomography aims to simultaneously reconstruct the full gate set using the maximum likelihood method~\cite{nielsen_gate_2021}. 
By measuring the outcomes prescribed by a list of gate sequences that acts to amplify errors on each gate, as shown in the bottom of Figure~\ref{fig:qgst}, one can run an optimization algorithm to find out all the parameterized process matrices within the gate set. 
It is precisely due to this `all in one' tomographic method that gate set tomography has the highest reconstruction accuracy versus traditional state tomography and process tomography, which are largely plagued by the problem of SPAM errors. 
However, the trade-off of gate set tomography is immediately obvious, in which way more computational resources have to be used to solve for this `simultaneous maximum likelihood' across all gate sequences.
This can be understood as maximizing the likelihood function in QGST is highly non-convex, in stark contrast to state and process tomography, where each observable probability is a linear function of the parameter~\cite{nielsen_gate_2021}. 
Based on this observation, using deep learning techniques to capture complex non-linear relationships would be a natural choice.

\subsubsection{Super operator formalism} 

Similar to the typical Dirac notation in Hilbert space, where a row vector is
represented by a bra  $\left\langle a\right|$, and column vector by a ket  $\left|b\right\rangle $, we denote superbra as $\langle\langle A|$ and superket as   $|B\rangle\rangle$. 
In quantum tomography settings, this conveniently maps a quantum state $\rho$ in the form of a $d\times d$ density matrix in the $d$-dimensional Hilbert space into a complex $d^2$-dimensional vector in Hilbert-Schmidt space, with the inner product defined as $\langle\langle A|B\rangle\rangle =\mathit{Tr}(A^{\dag{}}B)$.


In this paper, we use the Pauli Transfer Matrix (PTM) as our super-operator representation, as it is a popular choice in quantum tomography.
The PTM basis $\{B_i\}$ in Hilbert-Schmidt space has the following properties:
\begin{enumerate}
\item Hermiticity:  $B_i=B_i^{\dag}$
\item Orthonormality:  $\mathit{Tr}\left(B_iB_j\right)=\delta _{\mathit{ij}}$
\item Traceless for  $i>0$:  $B_0=I/\sqrt d$ \ and  $\mathit{Tr}\left(B_i\right)=0,$ for $i>0$
\end{enumerate}

For a single-qubit, the normalized PTM basis would be 
$\{B_i\}=\{I/\sqrt d,\sigma_x/\sqrt2,\sigma_y/\sqrt2,\sigma_z/\sqrt2\}$. 
Due to this choice of basis, the PTM vector and super operator are always real. 

As an example, we write a single-qubit  $2\times 2$ density matrix $\rho $ as  $|\rho \rangle\rangle$, represented by a real  $4\times 1$ column vector, where each coefficient of  $|\rho \rangle\rangle$ can be found by taking the inner product  $\mathit{Tr}\left(B_i^{\dag{}}\rho \right)$. 

\begin{equation*}
|\rho \text{⟩}\text{⟩}=\left[\begin{matrix}\mathit{Tr}\left(B_0^{\dag{}}\rho
\right)\\\mathit{Tr}\left(B_1^{\dag{}}\rho \right)\\\mathit{Tr}\left(B_2^{\dag{}}\rho
\right)\\\mathit{Tr}\left(B_3^{\dag{}}\rho \right)\end{matrix}\right]
\end{equation*}

To find the measurement probability of  $|\rho\rangle\rangle $ projecting onto the computational basis $\{\left|0\right\rangle ,\left|1\right\rangle\}$ we can perform the
standard dot product. 
First, we write the projectors as row vectors,
\begin{equation*}
|0\rangle\langle0|\ \mapsto\langle\langle E_0|=\ (1/\sqrt2,0,\ 0,\ 1/\sqrt2)
\end{equation*}
\begin{equation*}
|1\rangle\langle1|\ \mapsto\langle\langle E_1|=\ (1/\sqrt2,0,\ 0,\ -1/\sqrt2)
\end{equation*}

And, then, a standard dot product (or trace) obtains the measurement probabilities  $p_0, p_1$ of getting $0$ and $1$,
$$p_0=\left\langle \left\langle E_0\left|\rho \right.\right\rangle \right\rangle =\mathit{Tr}\left(E_0\rho \right)$$
$$p_1=\left\langle \left\langle E_1\left|\rho \right.\right\rangle \right\rangle =\mathit{Tr}\left(E_1\rho \right)$$

Naturally, for any  $d^2$ quantum state vector, we have the  $d^2\times d^2$ super operator that describes a (noisy) quantum channel, which is not necessarily unitary and/or orthogonal. 
For any quantum operator  $\Lambda $, the PTM satisfies
\begin{equation*}
\left\langle \left\langle j\left|R_{\Lambda }\right|k\right\rangle \right\rangle =\mathit{Tr}\left(\sigma _j\Lambda
\left(\sigma _k\right)\right)=\left(R_{\Lambda }\right)_{\mathit{jk}}
\end{equation*}
where, applying a quantum operation/channel  $\Lambda $ to a quantum state  $|\rho \rangle \rangle$ is
represented as left-multiplying a matrix to a vector,

\begin{equation*}
|\Lambda(\rho)\rangle\rangle=\ R_\Lambda|\rho\rangle\rangle
\end{equation*}
For instance, the PTM of a single-qubit rotational gate along the x-axis by  $\pi/2$ is given by,
\begin{equation*}
R_X\left(\pi \text{/}2\right)=\left[\begin{matrix}1&0&0&0\\0&1&0&0\\0&0&0&-1\\0&0&1&0\end{matrix}\right]
\end{equation*}

Corresponding to a quantum operation  $\Lambda \left(\rho \right)=U\rho U^{\dag{}}$, $U$ is the unitary single
qubit operator.

\subsection{Tomography using deep neural networks}

The goal of tomography in any general setting is to learn a latent space $z$ that maximally matches with the observed data $x$. 
The nature of this latent space depends on different scenarios.
For instance, referring back to medical imaging, 
the dimension of $z$ (the reconstructed 2D/3D CT image) is the same or higher than the observed data $x$ (2D sinograms). 
Notably, both $z$ and $x$ reside in the pixel space that requires little to no transformation when passed to typical neural network models like convolutional neural network~(CNN)~\cite{gupta_cnn-based_2018},~\cite{clark_convolutional_2019},~\cite{kang_deep_2017} or
diffusion model~\cite{huang_one_2022},~\cite{xia_sub-volume-based_2023},~\cite{xia_low-dose_2022}.

Quantum tomography is strikingly different as the $z$ and $x$ do not reside in the same space. 
For quantum state(process) tomography, the latent space $z$ is the density(process) matrix that takes the form of a square matrix, whereas the observed data $x$ refers to the measured counts or normalized probabilities 
from the quantum device, conditioned on a certain measurement and/or preparation operators. 

The same mismatch between latent space and observed space persists in gate set tomography, which prevents the direct implementation of typical deep generative models from image processing~\cite{ahmed_quantum_2021}.
Instead, an intermediate function has to be used in order to map the latent space to the observed space, namely, an analytical function that maps the neural network output to the expected probabilities under supervised learning.

The advancement of neural network models and computer hardware in the last
decade have brought forth numerous novel applications in the industry, such as autonomous robotics via reinforcement
learning~\cite{gu_deep_2017}, image generation~\cite{rombach_high-resolution_2022} via diffusion model, text generation via large language model~\cite{brown_language_2020} and so much
more. Riding on this trend, the quantum physics community has borrowed these techniques from the industry for quantum
tomography. Earlier work mainly focuses on using restricted Boltzmann machines for simple quantum state tomography tasks
that can be represented by pure states. Later works employ deep neural networks for more difficult tasks such as
general density matrix reconstruction~\cite{ahmed_quantum_2021} and process matrix construction~\cite{ahmed_gradient-descent_2023}
. The methods being used range from
simple feed-forward networks to more advanced models such as conditional generative adversarial networks and transformer
models. For instance, GAN demonstrated good convergence behavior in~\cite{ahmed_quantum_2021} when reconstructing a density matrix in the quantum state tomography setting, but this is only because the QST problem is linear in nature. It has been shown that GAN is susceptible to mode collapse~\cite{lala_evaluation_nodate}, which is particularly troubling when the data being trained on are multi-modal. In QGST, the problem that has to be solved is highly non-linear, with multi-modal data corresponding to multiple different gates that have to be estimated, as well as the underlying error parameters that represent those gates.

\section{\textsc{Ml4Qgst} for single-qubit systems} \label{s3}

Contrary to most existing publications that directly use deep neural networks to reconstruct the full density or process matrix in quantum state or process tomography settings, we aim to predict physical error parameters in gate set tomography instead and then use analytical functions to reconstruct the full process matrices afterward, as a way to ensure that the completely positive and trace preserving (CPTP) condition is met and remove the necessity of gauge fixing.

Furthermore, we alleviate the shortcomings of GANs by proposing a transformer model-based deep neural network, which excels in encoding and processing sequences~\cite{vaswani2017attention}, compared to standard convolution-based methods that are fundamentally limited by kernel sizes.
In addition, promising results have already been shown in quantum state tomography setting
recently using transformer-based techniques~\cite{cha_attention-based_2021},~\cite{ma_tomography_2023}. 

\added{In order to build up a solid foundation, we first explain the rationale behind our \textsc{Ml4Qgst} implementation and the corresponding experiments for single-qubit systems. In Section~\ref{2-qubit-intro}, we then generalize \textsc{Ml4Qgst} to multi-qubit systems based on our findings for single-qubit cases.}

\subsection{Transformer model}

Since the invention of the transformer architecture~\cite{vaswani2017attention}, the world has been revolutionized by the success and capabilities that GPT-4~\cite{achiam2023gpt}
and other large language models (LLM) provide. Here we briefly explain what a transformer is, specifically
the encoder block that we will use in this paper.

\begin{figure}
    \centering
    \includegraphics[width=\linewidth]{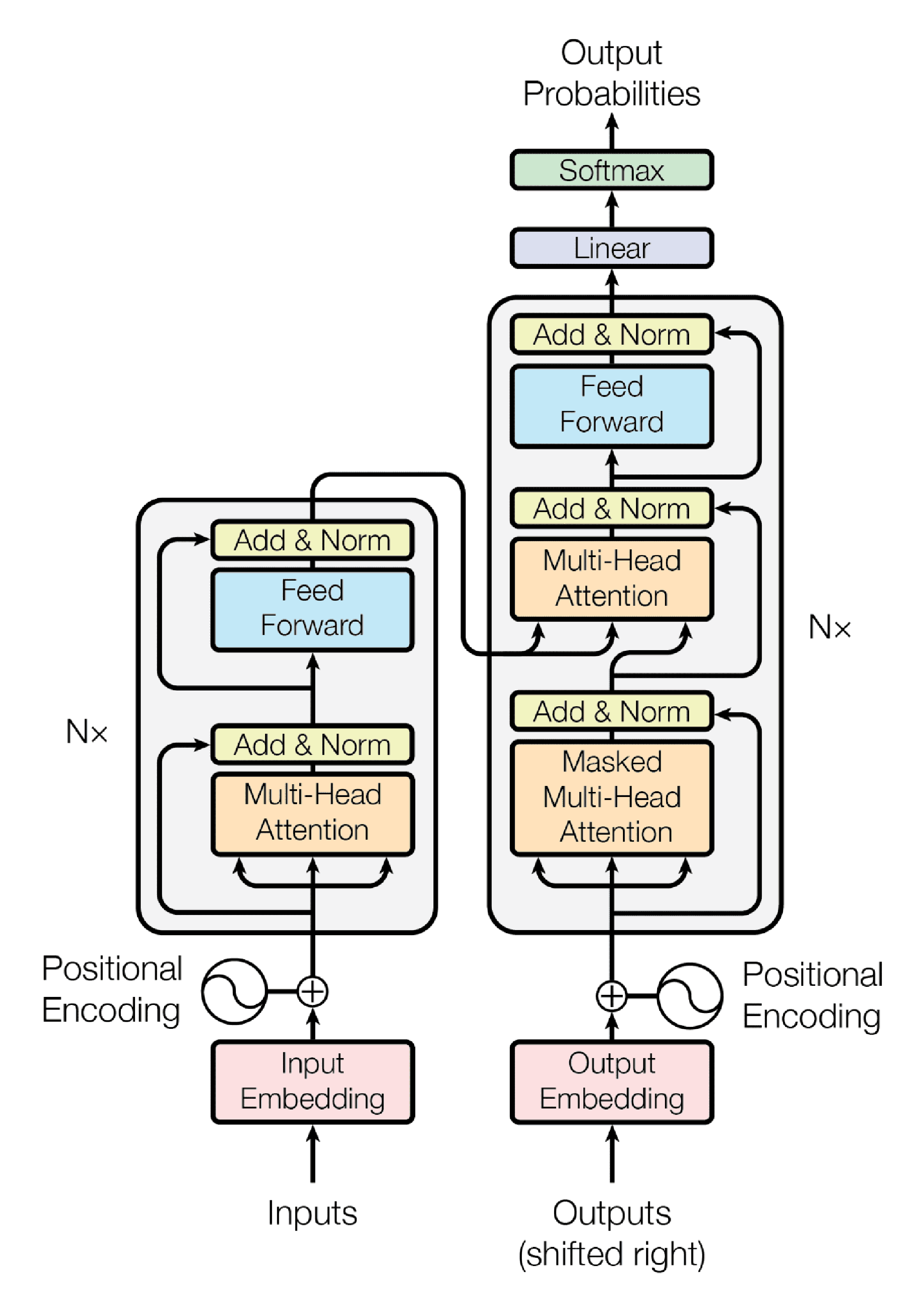}
    \caption{The Transformer model architecture with its encoder block (left) and decoder block (right) \cite{vaswani2017attention}.}
    \label{fig:transformer}
\end{figure}

Figure~\ref{fig:transformer} shows the complete transformer architecture that is used for text generation in the original implementation, and
it can be further divided into an encoder (left) and a decoder block (left). Within each block, the main component that
empowers the transformer is the multi-head attention layer (see orange-colored rectangle). As its name implies, the attention layer's goal is to
`pay attention' to the sequences, elements, or structures of the input data,
similar to what humans do. Mathematically, this is done by constructing arrays (query, key, value) and performing a dot product between query and key to obtain an attention score matrix. Afterward, a softmax operation is applied to the attention score matrix to obtain a new matrix corresponding to attention weights(probabilities). Finally, this attention weight matrix is multiplied with value, yielding a new weighted value output. The attention mechanism is most commonly seen in LLMs to process sequences of input text or, alternatively, focus on the underlying structure of images in computer vision~\cite{dosovitskiy_image_2021}.

In our model, we only make use of the encoder block to encode input data and subsequently use a simple feed-forward network for regression. We skip the transformer decoder commonly used in natural language processing, as our goal is not to generate new sequences but to predict error parameters.
This model encodes the gate sequence data naturally, similar to text encoding which has been widely adopted in the industry. 
We make use of self/cross attention mechanisms, which aim to focus on the information (process matrix) of each individual gate and the inter-relationship between gate sequences and normalized probabilities, respectively. 
By aggregating all the information through the transformer pipeline, we estimate the error parameters from the gate set tomography experiment.

\begin{figure}
    \centering
    \includegraphics[width=0.8\linewidth]{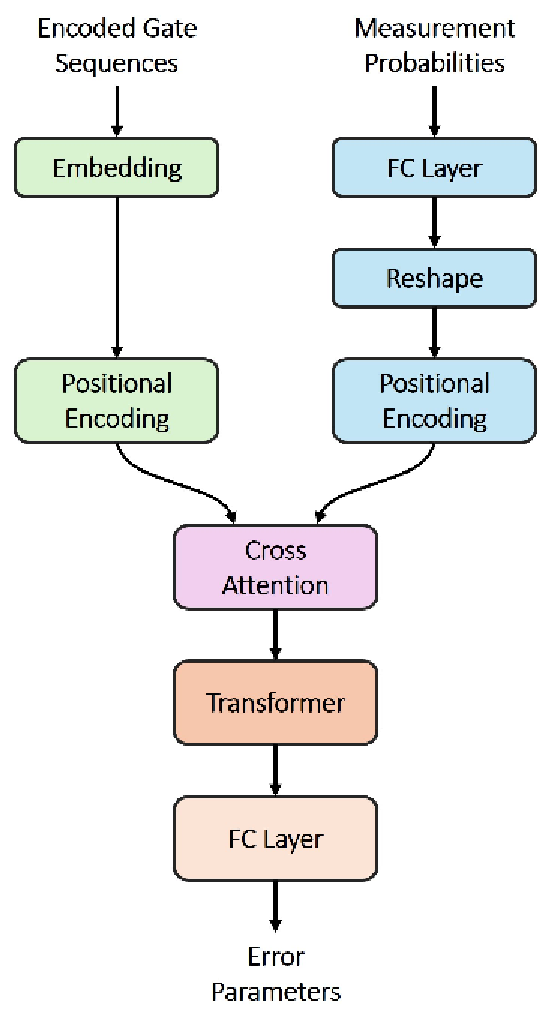}
    \caption{A schematic overview of \textsc{Ml4Qgst} \added{for single-qubit systems}, our transformed-based neural network architecture for GST.}
    \label{fig:NN_gst}
\end{figure}

The remainder of the section describes the main components of our \textsc{Ml4Qgst} model implementation for single-qubit systems.
Its components consist of 1) separate embedding layers for both
integer-encoded gate sequences and normalized probabilities, 2) separate positional encoding for both gate sequences and
normalized probabilities, 3) cross attention layer to aggregate information from two branches, 4) transformer block to
encode the aggregated information, 5) fully connected layer to output physical error parameters. A schematic of the
neural network architecture is shown in Figure~\ref{fig:NN_gst}.

\subsection{Embedding gate sequences and normalized probabilities} \label{Embedding gate sequences and normalized probabilities}

In gate set tomography, each gate sequence output measures counts for each computational basis and can be converted into normalized probabilities for each basis state.
By aggregating the information of
multiple pairs of gate sequences and normalized probabilities, one can extract the information of the process matrix of
each gate used. The gate sequences are first preprocessed by integer encoding and zero padding, where each gate is
mapped to a unique integer, and the gate sequences are zero-padded to match the longest gate sequence in the dataset.
The encoded gate sequences are then passed to an embedding layer used in a typical transformer setting. Besides that, the normalized probabilities are passed to a fully connected layer and are subsequently reshaped to emulate the
effect of an embedding layer.

This way, both branches will have an extra learnable feature dimension that is ready to be processed by a transformer
block later on.

\subsection{Positional encoding}
We use the standard sine and cosine positional encoding for both embedded gate sequences and normalized probabilities,
that are flattened beforehand. As each individual pair of gate sequence and normalized probabilities yields little
tomographic information, we instead group multiple pairs together to increase the receptive field. Positional
encoding is then applied element-wise to this flattened embedded gate sequence and normalized probabilities at each
branch. 

\subsection{Cross attention layer}
After embedding and positional encoding at each branch, a cross-attention layer is used to process the relationship
between the grouped gate sequences and normalized probabilities. We choose cross-attention instead of simple
concatenation to avoid the vast data shape mismatch between gate sequences and normalized probabilities that can possibly
drown out the training signal.

\subsection{Transformer encoder block}
A standard multi-layer transformer encoder block is used to process the aggregated information. It includes typical components such as a multi-head attention layer, add \& norm layer, and feed-forward layer.

\subsection{Fully connected layers}
Finally, fully connected layers are used for regression after the transformer block, outputting predicted physical error
parameters.

\section{Single-qubit experiments} \label{s4}

We evaluate \textsc{Ml4Qgst} \added{for single-qubit scenarios} using the open-source Python package pyGSTi~\cite{nielsen_probing_2020}, with the capability of: 1) customizing the process matrix of each individual gate within a gate set, 2) selecting appropriate fiducials for the customized gate set, 3) generating appropriate gate sequences for the GST experiment, 4) simulating measured counts for the gate sequences.

\subsection{pyGSTi simulation settings}

\begin{figure}
    \centering
    \includegraphics[width=0.8\linewidth]{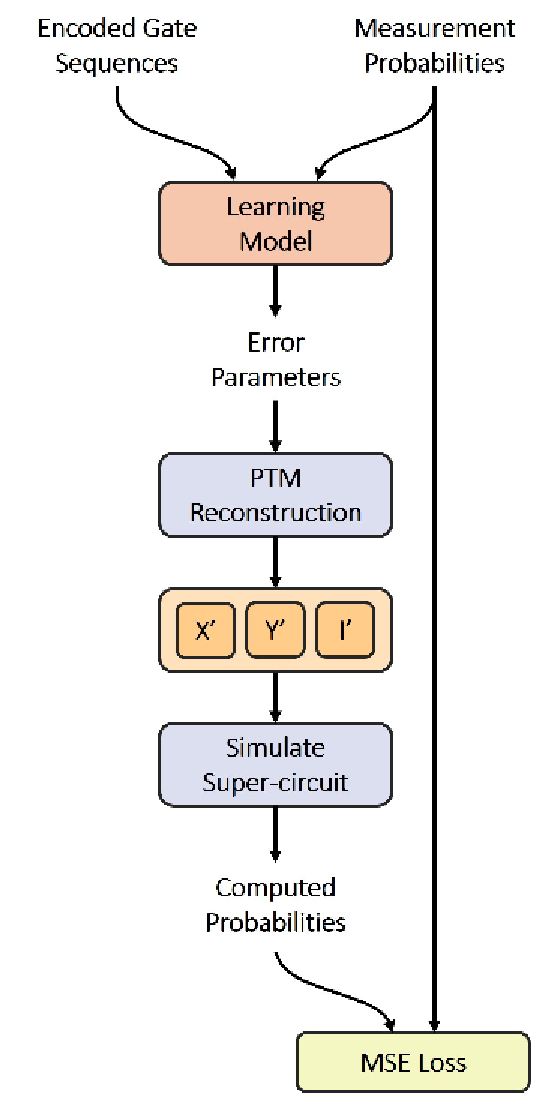}
    \caption{Overall data pipeline of \textsc{Ml4Qgst} from input to output and MSE loss computation for \added{single-qubit systems.}}
    \label{fig:NN_pipeline}
\end{figure}

The pyGSTi Python package uses Pauli Transfer Matrices (PTM) as the default process matrix throughout the QGST implementation. Here we replaced the built-in single-qubit XYI model with our custom PTM, specifically the X and Y rotational gates. These custom gates are parametrized with physical error parameters and in our case, the over-rotational angles and depolarizing errors. 
We then use the custom X and Y rotational gates, together with the built-in function, to find suitable fiducials, that is, a handful of short gate sequences that are used repeatedly and combinatorially to generate QGST experiments. After that, we run the built-in single-qubit XYI QGST experiment function to
generate gate sequences and simulated measured counts. The number of shots is set to 10k, and the maximum sequence length to 32, and the sampling error to binomial.
Figure~\ref{fig:NN_pipeline} shows the overall data pipeline from input to loss computation.

\subsection{Training details}

\subsubsection{Data grouping} \label{Data grouping}

As mentioned briefly in \ref{Embedding gate sequences and normalized probabilities}, the gate sequences generated by pyGSTi are converted from strings to unique integers and, subsequently, zero-padded to match the maximum length of a sequence in the dataset, whereas simulated measured counts are normalized into probabilities. After that, both gate sequences and probabilities datasets are divided into groups, specified by a hyperparameter `group\_size'.
In order to ensure all groups have the same number of elements, we choose to repeat the elements inside the last group instead of zero padding to preserve overall data quality.

\subsubsection{Curriculum learning}
Drawing inspiration from the way humans learn, curriculum learning aims to achieve better performance and faster convergence by starting with simpler or more fundamental examples and progressively introducing more complex ones \cite{bengio2009curriculum}, we conveniently call it stage training here, starting from easier to more difficult stages.
In the following, We make use of curriculum learning to further divide the whole dataset into parts, again specified by a hyperparameter `part\_size'. The dataset is sorted in ascending order based on the non-zero length of the gate sequences. This ensures that the model learns global features from shorter gate sequences in the beginning and then progressively fine-tunes predictions in later stages when it sees longer gate sequences. This learning methodology is similar to the algorithm implemented in the QGST paper~\cite{nielsen_gate_2021}, in which the authors iteratively add longer sequences, including the previously seen sequences in an accumulative way during optimization. We instead opt for a non-accumulative approach as in standard curriculum learning, and saving computational resources required. 

\subsubsection{Analytical PTM reconstruction}

Based on the predicted physical error parameters, namely the over-rotational angles and
depolarizing errors, we analytically reconstruct PTMs corresponding to the gates within the gate set.

\subsubsection{Computing loss}

As each grouped data outputs one set of physical error parameters prediction, it also has its own set of reconstructed PTMs. 
We compute probabilities analytically for all the gate sequences within a group, using the same set of reconstructed PTMs. This procedure is performed iteratively group by group within a particular stage set by curriculum learning. Finally, we compute the mean squared error loss between the ground-truth probabilities and the reconstructed probabilities. 

\section{Single-qubit results} \label{s5}
In the following, we will explain the choice of loss function in our experiment, analyze convergence behavior during neural network training and compare benchmarking results from different commonly used metrics. 

\subsection{Choice of loss function}

The QGST paper \cite{nielsen_gate_2021} uses two loss functions for long-sequence QGST optimization, the multinomial log-likelihood function $log\left(L\right)$ for a $m_s$ outcomes Bernoulli scheme, and the $\chi^2$ estimator.

$$log\left(L\right)=\sum_{s}\ log\left(L_s\right)=\sum_{s,\beta_s}\ N_sf_{s,\beta_s}log\left(p_{s,\beta_s}\right) $$

\begin{equation*}
\chi ^2=\sum _{s,\beta _s}\frac{N_s\left(p_{s,\beta _s}-f_{s,\beta _s}\right)^2}{p_{s,\beta _s}}
\end{equation*}
where $s$ denotes the index of a circuit, and let $m_s$ be the number of outcomes of $s$, $N_s$ the total number of times circuit $s$ was repeated, $N_{s,\beta _s}$ the number of times outcome $\beta _s$ was observed, $p_{s,\beta _s}$ the probability predicted by the model of getting outcome $\beta _s$ from circuit $s$, and $f_{s,\beta _s}=N_{s,\beta _s}/N_s$ is the corresponding observed frequency. \newline
The authors used the $\chi^2$ estimator as a proxy of $log (L)$ during optimization except for the last phase, as it is more computationally efficient. 
Then, $log (L)$ is used in the final phase to steer the estimate to comply with the true statistical derivation. 
Here, we further simplify the $\chi^2$ estimator to mean-squared error~(MSE) loss, which has also been used in simpler linear QGST settings. Alternatively, MSE loss can be seen from the perspective of reducing the likelihood function to a normal distribution by invoking the central limit theorem~\cite{greenbaum2015introduction}.

$$\mathit{loss}=\sum _{s,\beta _s}\frac{\left(p_{s,\beta _s}-f_{s,\beta _s}\right)^2}{\sigma _{s,\beta _s}^2}$$

where $\sigma _{s,\beta _s}^2$\ =  $p_{s,\beta _s}(1-p_{s,\beta _s})/N_s$ is the sampling variance in the
measurement.

\subsection{Convergence analysis}

In the following, we will explain the training trajectories of predicted error parameters by delving a little bit deeper into the technical implementation.

\begin{figure}[!htb]
    \centering
    \includegraphics[width=\linewidth]{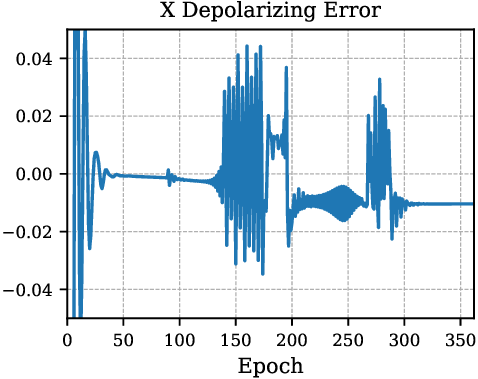}
    \caption{Training trajectory of the predicted X-gate depolarizing error with curriculum learning, tanh activation function is used at the final output layer to allow large gradient near zero, a subsequent absolute value function is added to ensure predicted value is between 0 and 1.}
    \label{fig:XerrCL}
\end{figure}

\begin{figure}[!htb]
    \centering
    \includegraphics[width=\linewidth]{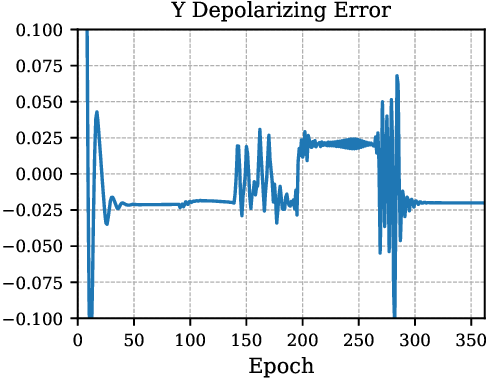}
    \caption{Training trajectory of the predicted Y-gate depolarizing error with curriculum learning, tanh activation function is used at the final output layer to allow large gradient near zero, a subsequent absolute value function is added to ensure predicted value is between 0 and 1.}
    \label{fig:YerrCL}
\end{figure}

\begin{figure}[!htb]
    \centering
    \includegraphics[width=\linewidth]{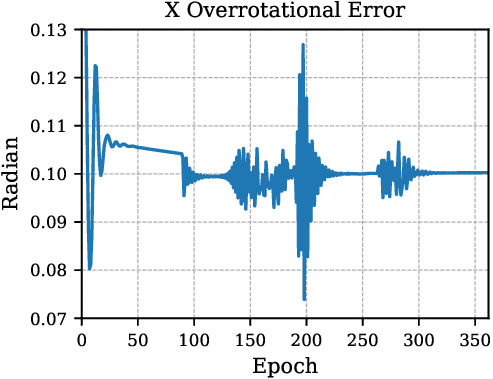}
    \caption{Training trajectory of the predicted X-gate over-rotational error with curriculum learning, tanh activation function is used at the final output layer to ensure the output value is between -1 and +1 (radian).}
    \label{fig:XrotCL}
\end{figure}

\begin{figure}[!htb]
    \centering
    \includegraphics[width=\linewidth]{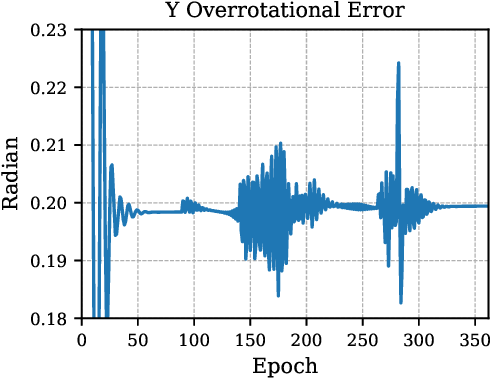}
    \caption{Training trajectory of the predicted Y-gate over-rotational error with curriculum learning, tanh activation function is used at the final output layer to ensure the output value is between -1 and +1 (radian).}
    \label{fig:YrotCL}
\end{figure}

Figures~\ref{fig:XerrCL},~\ref{fig:YerrCL} show the convergence behavior plots for depolarizing error and Figures~\ref{fig:XrotCL},~\ref{fig:YrotCL} for over rotational angle. 
Both types of plots contain the X-gate and Y-gate training trajectories, the $x$-axis always represents the number of training epochs and the $y$-axis indicates the depolarization amplitude and over rotational angle in radian respectively.
For both, the predicted depolarizing errors and the over-rotational angles, the predicted values exhibit oscillatory behavior at the beginning of each stage of curriculum learning, where an entirely new set of data was fed into the neural network for further training. This is indicated at epochs 90, 190, and 263, corresponding to the start of stages 2, 3, and 4. 

Because we use the tanh activation function at the neural network output layer and subsequently take absolute values in the custom training loop, the plots for depolarizing errors shown below will generally have the predicted values jumping between positive and negative. This is intended, as we want the predicted depolarizing error values to be close to and centered at zero, where the tanh activation function is the prime candidate. 

\begin{figure}[!htb]
    \centering
    \includegraphics[width=\linewidth]{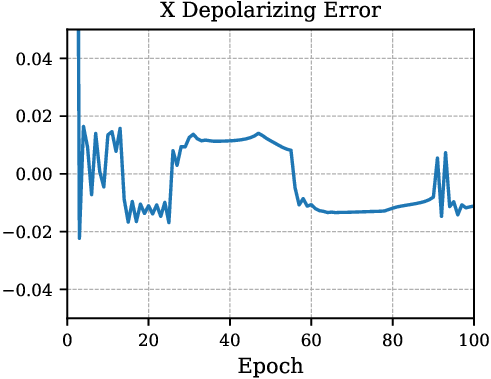}
    \caption{Training trajectory of the predicted X-gate depolarizing error without curriculum learning, tanh activation function is used at the final output layer to allow large gradient near zero, a subsequent absolute value function is added to ensure predicted value is between 0 and 1.}
    \label{fig:Xerr}
\end{figure}

\begin{figure}[!htb]
    \centering
    \includegraphics[width=\linewidth]{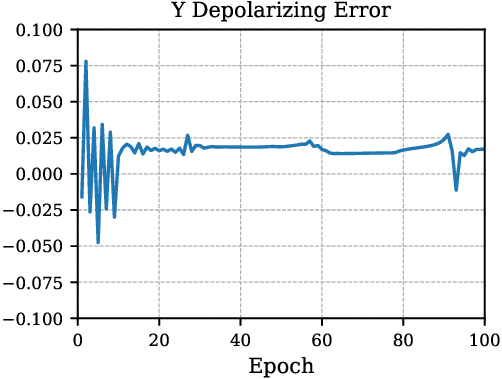}
    \caption{Training trajectory of the predicted Y-gate depolarizing error without curriculum learning, tanh activation function is used at the final output layer to allow large gradient near zero, a subsequent absolute value function is added to ensure predicted values is between 0 and 1.}
    \label{fig:Yerr}
\end{figure}

\begin{figure}[!htb]
    \centering
    \includegraphics[width=\linewidth]{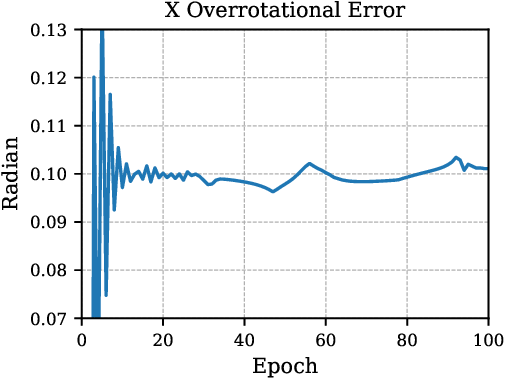}
    \caption{Training trajectory of the predicted X-gate over-rotational error without curriculum learning, tanh activation function is used at the final output layer to ensure the output value is between -1 and +1 (radian).}
    \label{fig:Xrot}
\end{figure}

\begin{figure}[!htb]
    \centering
    \includegraphics[width=\linewidth]{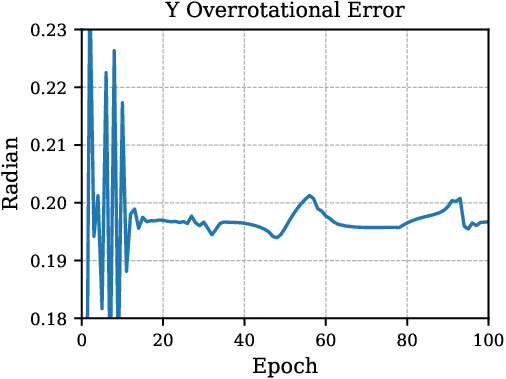}
    \caption{Training trajectory of the predicted Y-gate over-rotational error without curriculum learning, tanh activation function is used at the final output layer to ensure the output value is between -1 and +1 (radian).}
    \label{fig:Yrot}
\end{figure}

Additionally, we showed that without curriculum learning, the model fails to converge within the normalized number of epochs,
which is equal to the number of epochs for each stage in the curriculum learning. Figures~\ref{fig:Xerr},~\ref{fig:Yerr},~\ref{fig:Xrot},~\ref{fig:Yrot} show the convergence trajectories without curriculum learning. The empirical evidence here shows that curriculum learning, like the iterative optimization approach from the QGST paper, is indeed required for proper convergence.

\begin{table*}[!hbt]
\begin{tabular}{|l|l|l|l|p{2.4cm}|}
\hline
Loss function & With CL & Without CL & Ground-Truth & \% Error Ratio (W.O CL/  CL) \\ \hline
MSE (Training) & 1.9668e-05 (-0.08\%)  &  2.1339e-05 (-8.41\%) & 1.9683e-05 (0\%)  &  105.1 \\ \hline
KL divergence &  5.2119e-05 (-2.35\%) &  5.6215e-05 (-10.39\%) & 5.0923e-05 (0\%)  & 4.4  \\ \hline
$\chi^2$ estimator & 0.003118 (-2.06\%)  &  0.003380 (-10.64\%) & 0.003055 (0\%)  & 5.2  \\ \hline
-$log \left(L\right)$ & 16.11541 (-1.86e-5\%)  & 16.11554 (-9.93e-4\%)  & 16.11538 (0\%)  & 53.4  \\ \hline
\end{tabular}
\caption{Benchmarking for different loss functions: MSE, KL divergence, $\chi^2$ estimator, -log $L$, with and without curriculum learning.}
\label{table:losses}
\end{table*}

\subsection{Benchmarking}
\begin{figure*}[ht]
    \centering
    \includegraphics[width=0.85\linewidth]{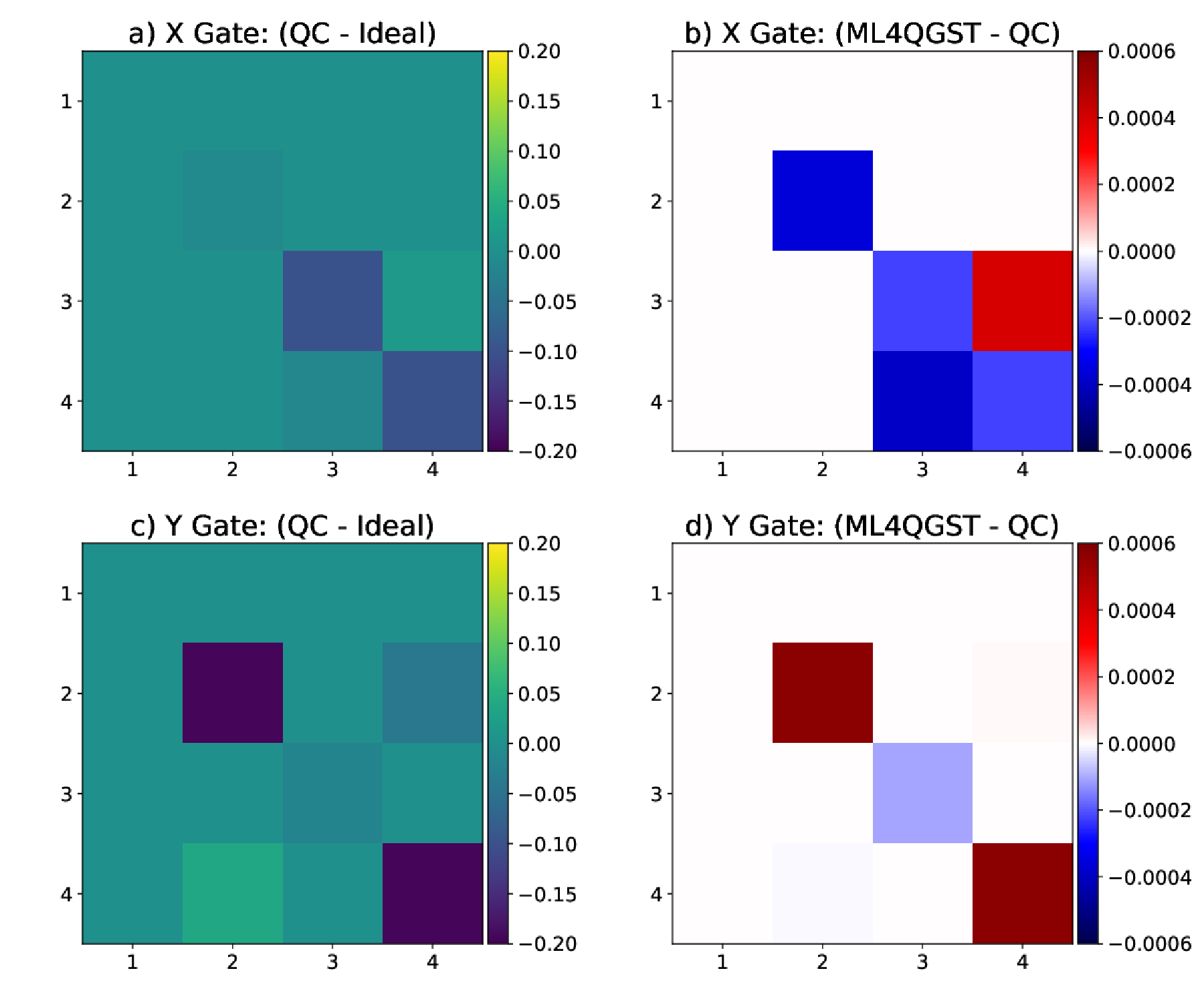}
    \caption{(a,b) X and Y gate PTM distance heatmaps between the ground-truth quantum computer (QC) and the Ideal operations, (c,d) X and Y gate PTM distance heatmaps between the ground-truth quantum computer (QC) and the \textsc{Ml4Qgst} predicted operations}
    \label{fig:hms}
\end{figure*}

To show that our predicted values are in good agreement with the ground-truth values from the simulation, we choose KL-divergence, $\chi ^2$ estimator, and full $log \left(L\right)$ function as a benchmark. We compare the benchmark results among three cases: ground-truth values, predicted values with curriculum learning (CL), and predicted values without curriculum learning, as shown in Table~\ref{table:losses}. The zero reference point for percentage error is set to ground truth in the table.

We first look at MSE, KL divergence, and $\chi ^2$ estimator, which all are the functions that measure the distance between probability distributions. It is no surprise that MSE has the lowest percentage error, as it was used as the loss function during training. We can also see a consistent trend among the three benchmarks, where the results without CL are much worse than the ones with CL, verifying the necessity of CL during training. Besides looking at just the percentage error, the error ratio between the known bad (without CL) and good (with CL) fit can roughly tell us how large the training signal would be, whereas a bigger ratio usually refers to a larger training signal. It can be seen that MSE and $-log \left(L\right)$ functions have large error ratios, meanwhile KL divergence and $\chi ^2$ estimator yield small ratios, suggesting MSE and $-log \left(L\right)$ functions are more versatile for training purpose.

Finally, we show the process matrix distance heatmaps in Figure~\ref{fig:hms}, as a standard practice to visualize the estimated results, commonly used in quantum tomography. To show that the model is indeed estimating the error parameters/ process matrices sensibly, the PTM distance heatmaps between (QC - Ideal) and (\textsc{Ml4Qgst} - QC) are compared, where ideal refers to operations with no depolarizing error and over rotation, QC means ground-truth quantum computer (QC) operations and \textsc{Ml4Qgst} refers to our model predicted operations. Figure \ref{fig:hms}(a,b) tells us how bad the quantum computer behaves with respect to the ideal operations that we actually want, while Figure \ref{fig:hms} (c,d) informs us how well our \textsc{Ml4Qgst} estimates the ground-truth quantum computer operations. Comparing the X and Y gate distance heatmaps in Figure~\ref{fig:hms} (c,d), we can roughly see the over-rotation angle and depolarizing error estimation slightly overshoot, with the exception of Y gate undershooting the estimation of over-rotation angle. The percentage differences between the ground truth and the estimated values are 3.6558\%(0.5321\%) for X(Y) depolarizing error and 0.2615\%(-0.2850\%) for X(Y) over-rotational angle. This is largely comparable to pyGSTi's long gate sequence QGST results: -0.0341\%(0.2670\%) and 0.5659\%(-0.3705\%).

\section{Generalizing to multi-qubit systems} \label{2-qubit-intro}
\added{Building upon the framework established for single-qubit systems, we further extend \textsc{Ml4Qgst} to handle multi-qubit scenarios, which are essential for scaling quantum processors and performing more complex quantum computations. Multi-qubit systems introduce significant additional challenges, including entanglement, cross-talk between qubits, and a combinatorial increase in the number of gate sequences required for comprehensive tomography. These complexities necessitate modifications to our original model to ensure accurate and efficient parameter estimation for higher number of qubits. In the following, we will introduce a revamped neural network architecture that is generalizable to multi-qubit systems. We then demonstrate this new model based on 2-qubit QGST data.}

\subsection{Modifications for multi-qubit systems}
\begin{figure}
    \centering
    \includegraphics[width=1\linewidth]{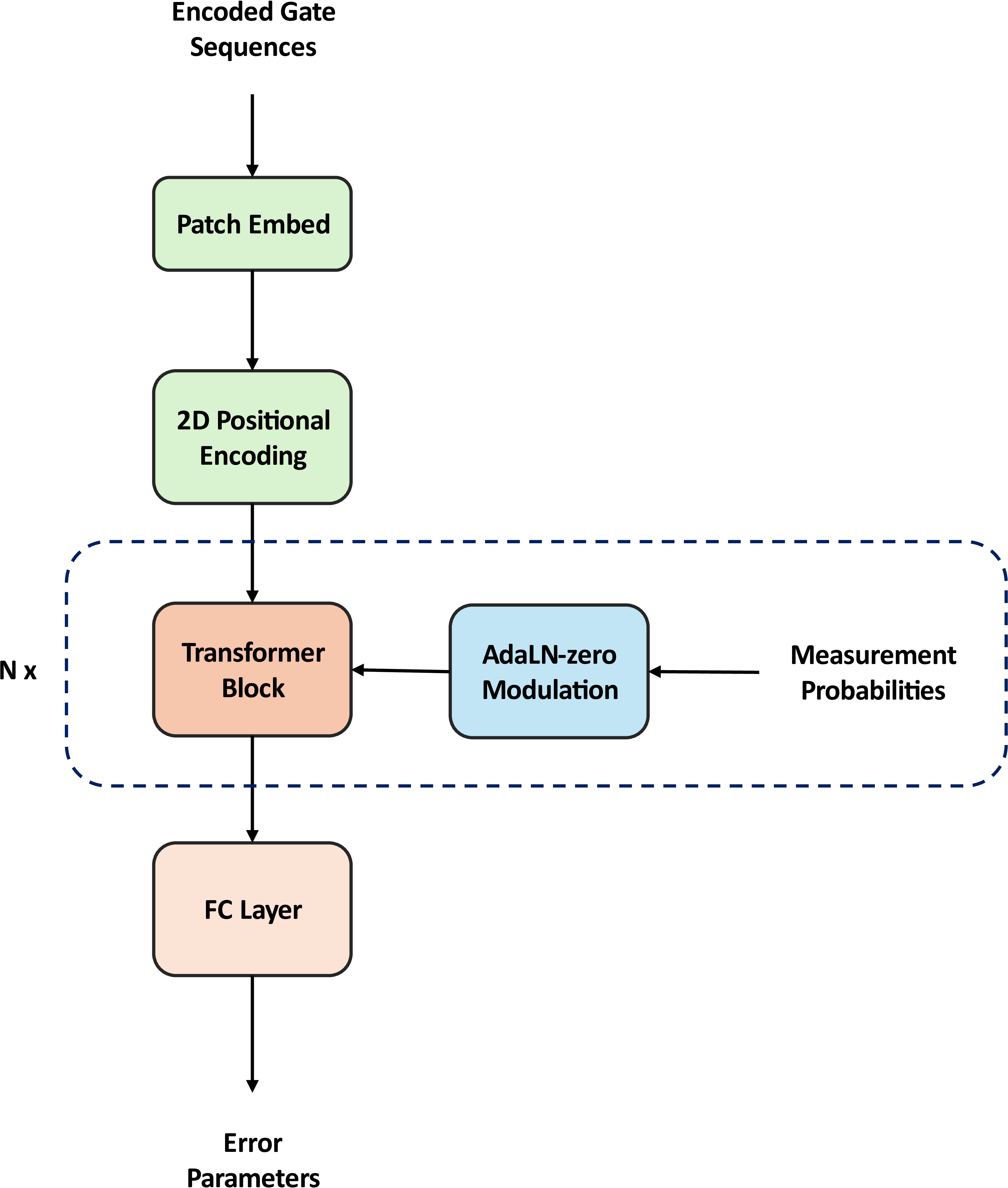}
    \caption{A schematic overview of our revamped neural network model for multi-qubit QGST, replacing 1D transformer with vision transformer and cross attention with adaLN-zero modulation.}
    \label{fig:NN_2q_gst}
\end{figure}

\added{We summarize the key modifications introduced in moving from a single-qubit model to a scalable multi-qubit model as follows:}
\begin{enumerate}
    \item \added{The architecture changes from a standard 1D transformer to a vision transformer, enabling efficient processing of multi-dimensional input data.}
    \item \added{The data grouping technique is extended by incorporating 2D sinusoidal positional encoding, which is better suited for representing spatial relationships in multi-qubit systems.}
    \item \added{The single time cross-attention layer used in the previous model is replaced by adaptive layer normalization (adaLN) with zero-modulation, applied at each transformer block. This change enhances the model’s ability to capture fine-grained interactions in the data.}
    \item \added{The loss function is changed from mean squared error (MSE) to Kullback–Leibler (KL) divergence loss, improving the model’s performance in tasks involving probabilistic distributions.}
\end{enumerate}
\added{Figure \ref{fig:NN_2q_gst} illustrates the revamped neural network architecture, which is specifically designed to be scalable for multi-qubit Quantum Gate Set Tomography (QGST) tasks.}

\subsubsection{Vision transformer}
\added{The most important modification of the model is migrating from typical 1D sequence transformer architecture to a vision transformer \cite{dosovitskiy2020image}. To recall, the $x$ and $y$ axis from the tokenized circuits represents gate sequence in time and qubit index, respectively. When increasing qubit number, the dimension along in $y$ axis increases,  effectively forming a 2D array. We can therefore treat any 2D array using the vision transformer. In contrary to standard transformer that can only handle 1D sequence data, vision transformer utilizes patch embedding to handle 2D arrays like natural images. During patch embedding, 2D arrays that have height $H$ and width $W$ are patchified into $H/p \times W/p$ patches, denoted by $p \times p$ patch size. These patches are then flattened and embedded to higher dimensional space, and subsequently fed into standard transformer block. Due to this architecture of vision transformer, it can be easily generalized to multi-qubit cases by simply altering the patch size. In our 2-qubit QGST experiment, we set $p$ = 2. Intuitively, setting $p$ equals to the number of qubits can preserve the global relationship along the qubit index axis, while setting $p$ smaller than the number of qubits could possibly introduce local bias. As for square patches, $p$ = 2 is the only possible option for the two qubit case. We leave the tuning of hyper-parameter $p$ in larger qubit cases for future work.}

\subsubsection{Data grouping and 2D positional encoding}
\added{We reuse the same data grouping technique as described in section \ref{Data grouping}, while introducing new 2D sinusoidal positional encoding to accommodate 2D arrays. We stack gate sequences within a group along the qubit index axis as before. By incorporating 2D positional encoding, we can now explicitly capture positional relationship of different gate sequences in a group.  
}

\subsubsection{AdaLN-zero conditioning modulation}
\added{Our previous single-qubit model only utilizes cross attention conditioning between the normalized probabilities and gate sequences before transformer blocks. This limits how the normalized probabilities affect the overall gradient flow during training. In our new model, instead of cross attention, we use adaLN-zero modulation to inject normalized probabilities conditioning at every transformer block. We observe superior training stability and convergence behavior versus simple cross attention.}

\subsubsection{KL divergence loss}
\added{Previously, the mean squared error (MSE) loss function was utilized due to its simplicity. In this study, we adopt the Kullback–Leibler (KL) divergence as the loss function to explore its effectiveness, as two-qubit QGST typically requires approximately ten times more circuits compared to the single-qubit case. To conserve computational resources, we reduce the number of shots per circuit from 10,000 to 1,000. Remarkably, we do not notice a significant drop in precision.}

\section{Two-qubit experiment} \label{2-qubit-experiment}
\added{In this section, we demonstrate our new model using 2-qubit QGST data, using the standard model pack smq2QXYICPHASE in pyGSTi~\cite{nielsen_probing_2020} to obtain fiducials, germs and a full QGST experiment. We then numerically simulate the full QGST experiment by employing a custom error model. Similar to the single-qubit case, our custom error parameters are depolarizing error and over-rotational error for three gates \{X, Y, CPHASE (Controlled-Phase)\}. We set the maximum depth of the circuit to 16 and 1,000 shots per sequence. In the following, we show the effectiveness of our model with two sets of 2-qubit QGST training data, with different strength in overrotational errors. At the end of this section, we will also illustrate the idea of transfer learning: fine-tuning an already trained model with a new set of training data to save computational resources.}

\subsection{Two-qubit convergence analysis}
\begin{figure}[!htb]
    \centering
    \includegraphics[width=\linewidth]{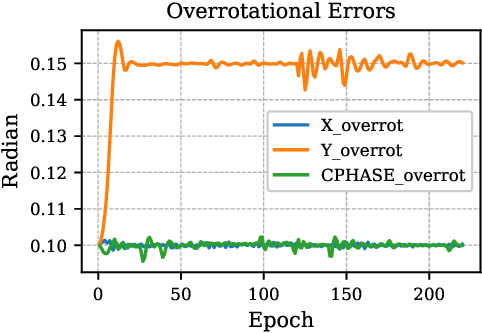}
    \caption{(Dataset 1) 2-qubit QGST with large rotational errors: Training trajectory of the predicted \{X, Y , CPHASE\} gates in over-rotational errors with curriculum learning. Tanh activation function is used at output layer.}
    \label{fig:large_error_overrot_err}
\end{figure}

\begin{figure}[!htb]
    \centering
    \includegraphics[width=\linewidth]{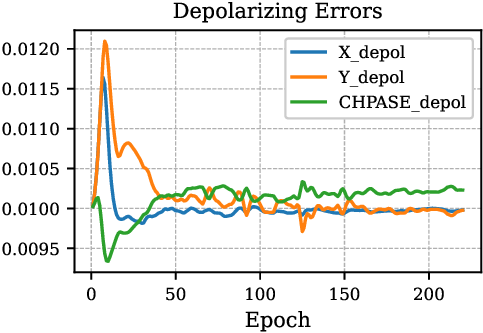}
    \caption{(Dataset 1) 2-qubit QGST with large rotational errors: Training trajectory of the predicted \{X, Y , CPHASE\} gates in depolarizing errors with curriculum learning. Sigmoid activation is used at output layer.}
    \label{fig:large_error_depol_err}
\end{figure}

\added{In the following, we show the training trajectories for 2-qubit QGST experiment. We start focusing on the dataset with large rotational errors, denoted as dataset 1. Figure \ref{fig:large_error_overrot_err} represents the training trajectory of the predicted \{X, Y , CPHASE\} gates in over-rotational errors with curriculum learning. We divided the 2-qubit QGST dataset in to three parts in ascending order of circuit depth for curriculum learning, and manually set training epochs to 60, 60, 100 for these three parts. Compared to the 1-qubit model, our 2-qubit model shows superior training stability and convergences quickly. We still use the tanh activation function as the output layer for over-rotational errors predictions, but without an extra absolute value function to restrict the output to be positive as for the 1-qubit model. Nonetheless, all the predicted values lie on the positive range without going into negative in later training stages. We attribute this superior training stability to adaLN-zero layer injection at each transformer block, where conditional normalized probabilities can guide the gradient flow much more efficiently. Figure \ref{fig:large_error_depol_err} also shows similar stable training trajectories for depolarizing errors. It can be observed that the convergence of depolarizing errors is slower than that of over-rotational errors, indicated by larger fluctuations before epoch 60 (end of part 1), which is expected as it requires longer circuits to accumulate large enough error signals. The predicted(ground-truth) over-rotational errors values for \{X, Y , CPHASE\} are 0.09963(0.1), 0.1505(0.15), 0.09938(0.1) respectively, and for depolarizing errors, the values are 0.009967(0.01), 0.009977(0.01), 0.01019(0.01) respectively.}

\begin{figure}[!htb]
    \centering
    \includegraphics[width=\linewidth]{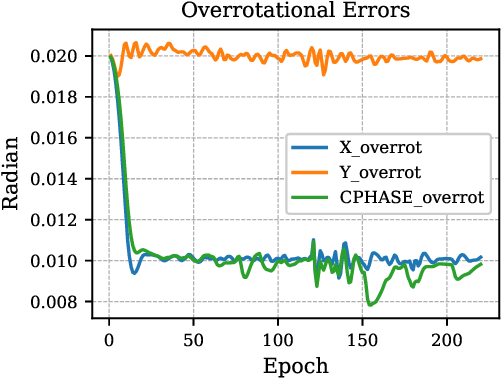}
    \caption{(Dataset 2) 2-qubit QGST with small rotational errors: Training trajectory of the predicted \{X, Y , CPHASE\} gates in over-rotational errors with curriculum learning. Tanh activation function is used at output layer.}
    \label{fig:small_error_overrot_err}
\end{figure}

\begin{figure}[!htb]
    \centering
    \includegraphics[width=\linewidth]{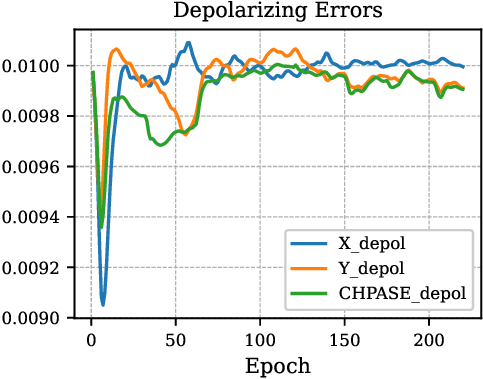}
    \caption{(Dataset 2) 2-qubit QGST with small rotational errors: Training trajectory of the predicted \{X, Y , CPHASE\} gates in depolarizing errors with curriculum learning. Sigmoid activation function is used at output layer.}
    \label{fig:small_error_depol_err}
\end{figure}

\added{Next, we demonstrate training trajectories for the dataset with small rotational errors, an order of magnitude smaller than the previous one, denoted as dataset 2. Figure \ref{fig:small_error_overrot_err}, \ref{fig:small_error_depol_err} show the training trajectories for over-rotational and depolarizing errors respectively, both indicate stable and smooth convergence behavior, just as the dataset with large over-rotational errors. The predicted(ground-truth) over-rotational errors values for \{X, Y , CPHASE\} are 0.01010(0.01), 0.01996(0.02), 0.009837(0.01) respectively, and for depolarizing errors, the values are 0.01001(0.01), 0.009939(0.01), 0.009938(0.01) respectively. Table \ref{table:2q-losses-dataset-1} and \ref{table:2q-losses-dataset-2} summarize all the predicted values versus ground-truth, and includes percentage errors for comparison. Most of the prediction have high accuracy of <1\% error, showing the model can perform accurate QGST analysis.}

\begin{table*}[!hbt]
\begin{tabular}{|l|l|l|p{2.4cm}|}
\hline
Gates & Dataset 1: Over-rotational Error & Dataset 1: Depolarizing Error & Percentage Error\\ \hline
X & 0.09963(0.1)  & 0.009967(0.01) & -0.3748, -0.3292   \\ \hline
Y &  0.1505(0.15) &   0.009977(0.01) & 0.3547, -0.2327 \\ \hline
CPHASE & 0.09938(0.1)  &  0.01019(0.01) & -0.6242, 1.9403  \\ \hline
\end{tabular}
\caption{Summary for the predicted(ground-truth) error values of gates, including percentage errors in dataset 1 (large over-rotational errors).}
\label{table:2q-losses-dataset-1}
\end{table*}

\begin{table*}[!hbt]
\begin{tabular}{|l|l|l|p{2.4cm}|}
\hline
Gates & Dataset 2: Over-rotational Error & Dataset 2: Depolarizing Error & Percentage Error \\ \hline
X & 0.01010(0.01)  & 0.01001(0.01)& 0.9595, 0.1367   \\ \hline
Y &  0.01996(0.02) &  0.009939(0.01) & -0.2130, -0.6097   \\ \hline
CPHASE & 0.009837(0.01)  &  0.009938(0.01) & -1.6343, -0.6241  \\ \hline
\end{tabular}
\caption{Summary for the predicted(ground-truth) error values of gates, including percentage errors in dataset 2 (small over-rotational errors).}
\label{table:2q-losses-dataset-2}
\end{table*}

\subsection{Transfer learning}
\begin{figure}[!htb]
    \centering
    \includegraphics[width=\linewidth]{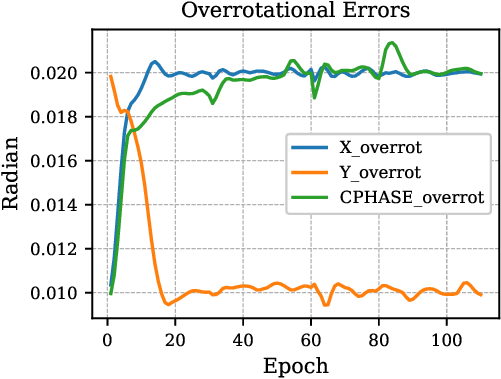}
    \caption{(Dataset 3) 2-qubit QGST with transfer learning: Training trajectory of the predicted \{X, Y , CPHASE\} gates in over-rotational errors with curriculum learning. Tanh activation function is used at output layer.}
    \label{fig:small_error_transfer_learning_overrot_err}
\end{figure}

\begin{figure}[!htb]
    \centering
    \includegraphics[width=\linewidth]{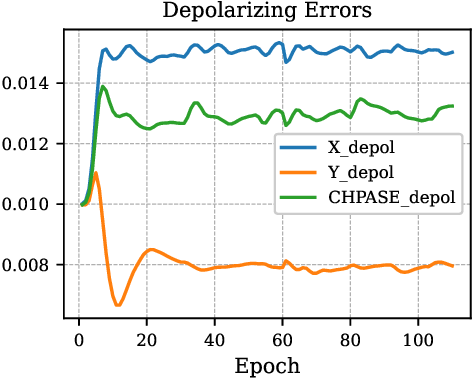}
    \caption{(Dataset 3) 2-qubit QGST with transfer learning: Training trajectory of the predicted \{X, Y , CPHASE\} gates in depolarizing errors with curriculum learning. Sigmoid activation function is used at output layer.}
    \label{fig:small_error_transfer_learning_depol_err}
\end{figure}

\added{In addition to the main 2-qubit QGST experiment with two different datasets, we show that the model can be fine-tuned to predict new error parameters, given a new set of data via transfer learning with reduced number of training iterations, denoted as dataset 3. In the main experiment, we set the training epochs to 60, 60, 120. Here in transfer learning, we set all the epochs to half of the original values to 30, 30, 60, so as to demonstrate this idea. We use a pre-trained model trained with dataset 2, and then fine-tune it using a new dataset with new error parameters. Figure \ref{fig:small_error_transfer_learning_overrot_err} and \ref{fig:small_error_transfer_learning_depol_err} show the training trajectories for over-rotational and depolarizing error respectively, which conforms well with the general trend in the main experiment. The predicted(ground-truth) over-rotational errors values for \{X, Y , CPHASE\} are 0.01998(0.02),  0.009931(0.01), 0.02012(0.02) respectively, and for depolarizing errors, the values are 0.01504(0.015), 0.007937(0.008), 0.01277(0.013) respectively. Table \ref{table:2q-losses-transfer-learning} shows that again most of the predicted values have accuracy of <1\% error.}

\begin{table*}[!hbt]
\begin{tabular}{|l|l|l|p{2.4cm}|}
\hline
Gates & Dataset 3: Over-rotational Error & Dataset 3: Depolarizing Error & Percentage Error \\ \hline
X & 0.01998(0.02) & 0.01504(0.015)& -0.1010,  0.2675 \\ \hline
Y &  0.009931(0.01)&  0.007937(0.008) & -0.6926, -0.7927  \\ \hline
CPHASE & 0.02012(0.02)  & 0.01277(0.013) & 0.6005, 3.8513  \\ \hline
\end{tabular}
\caption{Summary for the predicted(ground-truth) error values of gates, including percentage errors in dataset 3 (transfer learning).}
\label{table:2q-losses-transfer-learning}
\end{table*}

\section{Outlook} \label{Outlook}
In this section, we discuss possible use cases for \textsc{Ml4Qgst}, \added{focusing on bootstrapping errors and scalable tomographic reconstruction in multi-qubit systems.}

The most immediate and universal use case for a trained deep neural network model would be bootstrapping existing tomographic approaches, where an end user queries the trained model to obtain a list of fairly accurate predictions and subsequently uses this as an initial guess input to another mathematically rigorous traditional analytical or numerical model, vastly reduces the computation resource and time required from traditional approaches, while still retaining the superior prediction accuracy, compared to using deep neural network alone.

The next use case would be scaling up gate set tomography to multi-qubit systems. Although not \added{explicitly} demonstrated in this proof-of-concept paper, the transformer has been proven for capturing long-range relationships in natural language processing, which, in turn, should work well when processing multi-qubit long sequence quantum circuits. To draw a parallel, in terms of encoding quantum circuits for specific use cases, the transformer will be a natural extension and improvement to existing CNN-based quantum circuit optimization \cite{fosel2021quantum}. \added{In addition, the vision transformer has already been widely used to process high resolution natural images, often with a size larger than 1024 $\times$ 1024. Hence, there are empirical evidence that, even for a large amount of qubits, using a group of tokenized circuit  with size: [(number of qubits $\times$ group size), circuit length], the vision transformer can still efficiently handle the data.} In essence, we welcome future research in the quantum community to make use of transformer models to process quantum circuit-related applications.

\section{Conclusion} \label{s6}

In this article, \added{we presented \textsc{Ml4Qgst}},  a transformer-based neural network model for quantum gate set tomography. \added{Specifically, we employ a 1D transformer architecture with simple cross-attention conditioning for single-qubit model and a vision transformer enhanced with adaLN-zero modulation for multi-qubit model.} \added{ By incorporating curriculum learning, the proposed neural networks demonstrate stable convergence, and the final estimation results agree with the ground truth values.} Our results are a proof of concept that demonstrates that deep neural network models can also be used in tackling difficult highly non-linear tomography problems like gate set tomography. We wish to further improve the efficiency and accuracy of the model for gate set tomography in the future by exploring different neural network architectures and via model fine-tuning. In particular, we believe that leveraging the success of first compressing data in latent space via an auto-encoder that was recently used in diffusion model~\cite{rombach_high-resolution_2022} would greatly reduce the computational footprint for QGST. \replaced{}{QGST can be understood as mapping a large set of measurement data into a small subset of error parameters, where the idea of latent space would naturally come into the picture.}

\backmatter








\section*{Software Availability} \label{code_and_dataset}

The software developed for this project is available at: \href{https://github.com/QML-Group/ML4GST}{https://github.com/QML-Group/ML4GST}. 

\noindent The QGST data is generated via the pyGSTi software: \href{https://github.com/sandialabs/pyGSTi}{https://github.com/sandialabs/pyGSTi}.

\section*{Acknowledgments} \label{sec-acknowledgements}
KYY and AS would like to thank Tim Taminiau for useful discussions on the current experimental limits of device characterization. AS acknowledges funding from the Dutch Research Council (NWO). This research was partly supported by the EU through the H2024 QLSI2 project and partly sponsored by the Army Research Office under Award Number: W911NF-23-1-0110. The views and conclusions contained in this document are those of the authors and should not be interpreted as representing the official policies, either expressed or implied, of the Army Research Office or the U.S. Government. The U.S. Government is authorized to reproduce and distribute reprints for Government purposes notwithstanding any copyright notation herein.

\bibliographystyle{bst/sn-mathphys}
\bibliography{sn-bibliography}

\end{document}